\documentclass[aps,pra,superscriptaddress, twocolumn]{revtex4}
\usepackage[intlimits]{amsmath}
\usepackage{amsfonts, amssymb}
\usepackage{psfrag}
\usepackage{graphics}
\usepackage{subfigure}
\usepackage[usenames]{color}
\usepackage{multirow}
\usepackage{braket}
\usepackage{epstopdf}
\usepackage{tikz}

\usepackage{ifpdf}

\ifpdf
\usepackage{epstopdf}
\usepackage[pdftex]{hyperref}
\else
\usepackage[hypertex,ps2pdf]{hyperref}
\fi

\begin{document}

\title{\Large{The $\hbar\rightarrow 0$ Limit of the Entanglement Entropy}}
\author{G. Mussardo}
\affiliation{SISSA and INFN, Sezione di Trieste,
Italy}
\author{J. Viti}
\affiliation{International Institute of Physics \& ECT, UFRN, Campos Universitário, Lagoa Nova, Natal, 59078-970, Brazil}
 
.

\vspace{8mm}

\begin{abstract}
Entangled quantum states share properties that do not have classical analogs, in particular, they show correlations that can violate Bell inequalities. It is therefore an interesting question to see what happens to entanglement measures -- such as the entanglement entropy for a pure state -- taking the semiclassical limit,  where the na\"{i}ve expectation is that they may become singular or zero. This conclusion is however incorrect. In this paper, we determine the $\hbar\rightarrow 0$ limit of the
bipartite entanglement entropy for a one-dimensional system of $N$ quantum particles in an external potential and 
we explicitly show that this limit is finite. Moreover, if the particles are fermionic, we show that the 
$\hbar\rightarrow 0$ limit of the
bipartite entanglement entropy coincides with the Shannon entropy of $N$ bits.
\end{abstract}
\maketitle

\newpage 

\newpage
\vspace{5mm}

\section{Introduction}

The study of entanglement measures~\cite{Plenio}, which was firstly motivated by quantum information and computation~\cite{CN}, is nowadays central in many areas of theoretical physics. For a spatially extended quantum system in a pure state, a measure of entanglement that is invariant under local operations and classical communications~\cite{BBPS} is the Von Neumann entropy of the reduced density matrix, also known as {\em entanglement entropy}. This quantity has found extensive applications in both high energy~\cite{RT, N18} and condensed matter theory ~\cite{Fazio, ECP}.  

Entanglement~\cite{Horo, Islam, Yin} is rightly considered one of the key features of quantum mechanics that makes some of its predictions incompatible with any local classical theory~\cite{Bell2, Bell_review}.  By pushing this line of thought, one could expect that entanglement measures should be zero or singular when evaluated in a---properly defined---classical limit~\cite{Peres}.  This conclusion is however wrong, as we easily show by means of a simple example. 
\begin{itemize}
    \item 
Consider the quantum state
\begin{equation}
\label{state}|\Psi_{\eta}\rangle=\frac{1}{\sqrt{2}}(|\eta\rangle_A|0\rangle_B+|0\rangle_A|\eta\rangle_B),
\end{equation}

where $A$ and $B$ denote two spatially separated boxes and $$|\eta\rangle=\frac{1}{\sqrt{\eta!}}(a^{\dagger})^{\eta}|0\rangle$$ is the eigenstate of a harmonic oscillator with energy $$E_{\eta}=\hbar\omega(\eta+1/2), \,\,\,\eta=1,2,\dots$$ The state $|\Psi_{\eta}\rangle$ is entangled~\cite{Enk} and a bipartite entanglement entropy could be defined as  
\begin{equation}
\label{entropy_def}
S^{(\eta)}(A)=-\text{Tr}[\rho_A^{(\eta)}\log\rho_A^{(\eta)}]
\end{equation}
where $\rho_A^{(\eta)}=\text{Tr}_B|\Psi_{\eta}\rangle\langle\Psi_{\eta}|$. An elementary calculation gives $S^{(\eta)}(A)=\log 2$,  independently from the parameter $\eta$.

The value of $S^{(\eta)}(A)$ could be inferred by repeating the following experiment: the state $|\Psi_{\eta}\rangle$ is prepared by a certain device, and then the box $A$ is opened and a detector tests for the presence of a particle with energy $E_{\eta}$ inside the box. The Shannon entropy~\cite{Shannon, Witten_info} of the probability distribution to observe the particle inside $A$ is the entanglement entropy $S^{(\eta)}(A)$. Now  consider the same problem when $\eta\rightarrow\infty$. In this limit, for the correspondence principle, the wave function of the quantum state $|\eta\rangle$  can be approximated  with an arbitrary precision by the semiclassical wave function~\cite{Wheeler,Landau-3}.  The same experiment described above can be performed to recover the value $\lim_{\eta\rightarrow\infty}S^{(\eta)}(A)=\log 2$ for the entanglement entropy. However, in this case, the measurement outcomes could be explained without a knowledge of quantum mechanics. The observer can assume, for instance, that the experimental device injects with probability $1/2$ a classical particle inside the box $A$: if the particle is detected in box $A$, it will be not found in the box  $B$. Hence, the entanglement entropy of the state $|\Psi_{\eta}\rangle$ has trivially a finite classical limit which can be interpreted as the Shannon entropy of a bit. Similarly, the simple experiment which we just described,  fails to spot the difference between classical and quantum correlations of the state in Eq.~\eqref{state}. To this end,  it will be necessary to set up a different measurement protocol in the spirit of Ref.~\cite{Bell2}--(see also the discussion in Sec.~\ref{sec:classical}). 
\end{itemize}
In this paper, we  examine a similar problem for a one-dimensional quantum gas of $N$ particles in the presence of an external potential.
We still use a positive integer $\eta$ to label the energy quantum number of each particle and the classical limit is again defined by sending  $\eta\rightarrow\infty$ ~\cite{Wheeler}.
This condition implies that the classical action of each particle is much larger than $\hbar$ and therefore the limiting procedure is equivalent to the asymptotic expansion of the Schr\"odinger equation for $\hbar\rightarrow 0$~\cite{Landau-3}. The existence of the $\hbar\rightarrow 0$ limit, \textit{tout court}, is however subtle and might depend on the observable considered~\cite{Peres}. We note that the behavior of the entanglement entropy in the $\hbar\rightarrow 0$ limit  has been also discussed  in a time-dependent context~\cite{Angelo, Matzkin, Casati, Asplund, Lerose1}.

\begin{figure}[t]
\includegraphics{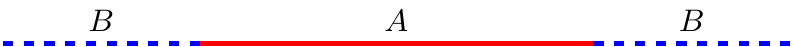}
\caption{Bipartition of the space adopted for the calculation of the entanglement entropy in the classical limit $\hbar\rightarrow 0$. In particular, we focus on the reduced density matrix $\rho_A$ and its Von Neumann entropy $S(A)=-\text{Tr}[\rho_A\log\rho_A]$.}
\label{fig_bip}
\end{figure}

The paper is organized as follows. In Sec.~\ref{sec:op}, we  introduce a quantum system of $N$  particles in an external potential. The full real line is partitioned into two intervals: the interval $A$ which is the one that will be observed and its complementary $B$ as in Fig.~\ref{fig_bip}. Given an $N$-particle state $|\Psi\rangle$, we study then the entanglement entropy calculated from the reduced density matrix $\rho_A=\text{Tr}_B|\Psi\rangle\langle\Psi|$. This setup is common for quantum chains and their continuum limit~\cite{W, Peschel2, Peschel, Eisert-h, Latorre, Korepin, Keeting,  CC, CastroCardy, Casini}.
In Sec.~\ref{sec:op} and Sec.~\ref{sec:classical} we initially focus our attention on the case $N=1$ and calculate the classical limit of the eigenvalues of the reduced density matrix.  In Sec.~\ref{sec:tp}, we study the $\hbar\rightarrow 0$ limit of the bipartite entanglement entropy for a two-particle state of bosons and fermions. At low energies, the entanglement properties of quasiparticle excitations  in quantum chains and their continuum limit have been investigated  in detail~\cite{Alba, Sierra,Berkovits, Alba2, Storms, Ares, Castro, Castro2, Castro3, Raja_Z, Raja}. For states with finite particle density we refer instead to~Refs.~\cite{Page, Grover, Rigol}.

Finally in Sec.~\ref{sec:classical} we discuss the general case of a  fermionic gas of $N$ particles  and we show that in the classical limit the subsystem entanglement entropy reduces to the Shannon entropy of $N$ bits. This analytic result extends the field theoretical calculation for free bosons on a ring of Refs.~\cite{Castro, Castro2} in the limit of large particle  momenta--(see also~Ref.~\cite{Piz}). 

Our conclusions can be found in Sec.~\ref{sec:conc}. The paper also contains two Appendices.

\section{Single particle entanglement entropy}
\label{sec:op}
Let us start our analysis with the simplest possible case, i.e. the study of the entanglement entropy coming from a pure state $| \Psi \rangle $ of a quantum particle living in a one-dimensional system in the presence of an external potential. Referring to Fig. \ref{fig_bip}, we will focus our attention on the degrees of freedom relative to the interval $A$ once we have integrated out those of the external intervals $B$. 

First of all, it is important to observe that in the first quantization formalism there is no notion of a reduced density matrix relative to a space interval $A\subset\mathbb R$, since in this case the Hilbert space cannot be factorized. Moreover, the first quantization formalism does not allow the possibility that the state of the system inside  region $A$ will be the vacuum, if the particle is detected outside. To overcome these difficulties, we reformulate the problem in the second quantization scheme. It is also convenient to define the system on a lattice with lattice-spacing $a$ and taking later the continuum limit $a\rightarrow 0$. Consider then the Hamiltonian 
\begin{equation}
\label{model}
 H=-\frac{\hbar^2}{2ma^2}\sum_{j\in\mathbb Z}(C^{\dagger}_j C_{j+1}+H.c.)+\sum_{j\in\mathbb Z} \left(V_j+\frac{\hbar^2}{ma^2}\right)C^{\dagger}_j C_j,
\end{equation}
in which we assume that the creation and annihilation operators $C_j$  and $C_j^{\dagger}$ are fermionic operators~\cite{Foot} of a particle of mass $m$.
A basis for the Hilbert space $\mathcal{H}$ is given by  the vectors   $\otimes_{j\in\mathbb Z}\{|0_j\rangle, |1_j\rangle\}$ where  $|1_j\rangle\equiv C^{\dagger}_j|0_j\rangle$, $C_j|0_j\rangle=O$. The vacuum state is defined by $|\Omega\rangle\equiv\otimes_{j\in\mathbb Z}|0_j\rangle$, therefore with a slight abuse of notation we also have $|1_j\rangle=C^{\dagger}_j|\Omega\rangle$.   The single particle eigenstates of Eq.~\eqref{model} can be then conveniently   labeled in terms of an integer number $\eta$ and written as  
\begin{equation}
\label{sp}
 |\Psi_{\eta}\rangle\,=\, \sum_{j\in\mathbb Z}\psi_j^{(\eta)}|1_j\rangle.
\end{equation}
The complex amplitudes $\psi_j^{(\eta)}$ in Eq.~\eqref{sp} solve the discrete Schr\"odinger equation
\begin{equation}
 -\frac{\hbar^2}{2m}
 \frac{
 \psi_{j+1}^{(\eta)} - 2\psi_j^{(\eta)} + \psi_{j-1}^{(\eta)}
 }
 {a^2} 
 + V_{j} \psi_j^{(\eta)} \, 
 = \, E_{\eta} \psi^{(\eta)}_{j}.
\end{equation}
The continuum limit $a\rightarrow 0$ is obtained by keeping $x=aj$ finite and further requiring
\begin{equation}
\label{cont_lim}
\psi_j^{(\eta)}/\sqrt{a}\rightarrow \psi^{(\eta)}(x), ~~C_j/\sqrt{a}\rightarrow C(x)
\end{equation}
where  $\psi^{(\eta)}(x)$ is the normalized wave function solution of the Schr\"odinger equation for a potential $V(x)$ and $C(x)$ is the annihilation operator in the continuum. 

For instance, when $V = 0$, for a particle living on the lattice of a one-dimensional box of length $L$, one has
\begin{equation}
 \psi_j^{(\eta)}=\mathcal{N}_{\eta}\sin \left(\frac{\pi \eta \,j }{L+1}\right) ~,~ \eta=1,2\dots
\end{equation}
where $\mathcal{N}_{\eta}$ a normalization constant while the corresponding energy eigenvalue is given by 
\begin{equation}
\label{eigenbox}
 E_{\eta}=-\frac{\hbar^2}{m a^2}\left[\cos\left(\frac{\pi \eta}{L+1}\right)-1\right] \,\,\,. 
\end{equation}
From Eqs.~\eqref{cont} and \eqref{eigenbox} it follows that  $E_\eta\stackrel{a\rightarrow 0}{\rightarrow}\frac{\hbar^2\pi^2 \eta^2}{2m\ell^2}$ with $\ell\equiv La$ as expected.

\vspace{3mm}
\noindent
\textit{Reduced density matrix---}Let us now consider a bipartition of our quantum system into  two spatial regions. The interval $A\subset\mathbb R$ made of $|A|$ sites and its complementary $B=\mathbb R\backslash A$ as in Fig.~\ref{fig_bip}. The full Hilbert space will be factorized as $\mathcal{H}=\mathcal{H}_A\otimes\mathcal{H}_B$ and, for fermionic particles, $\dim(\mathcal{H}_A)=2^{|A|}$.

The reduced density matrix relative to region $A$ is defined as $\rho_A^{(\eta)}\equiv\text{Tr}_B|\Psi_\eta\rangle\langle\Psi_\eta|$ and, after substituting Eq.~\eqref{sp}, it can be easily computed. The result is 
\begin{equation}
\label{reduced1pt}
 \rho_{A}^{(\eta)}=\sum_{r\not\in A}|\psi_r^{(\eta)}|^2 \Pi_{0}^{A}+\sum_{r,s\in A}\psi_r^{(\eta)}[\psi_s^{(\eta)}]^{*}|1_r\rangle\langle 1_s|.
\end{equation}
In Eq.~\eqref{reduced1pt} above, $\Pi_{0}^{A}$  is the projector on the zero-particle sector of the Hilbert space $\mathcal{H}_A$, namely $$\Pi_{0}^{A}=\otimes_{j\in A}|0_j\rangle\langle 0_j|\,\,\,,$$ while $\sum_{r,s\in A}|1_r\rangle\langle 1_s|$ is the projector on the one-particle sector. The two operators are orthogonal and, as a matrix acting on $\mathcal{H}_A$,  $\rho_A^{(\eta)}$ is then block diagonal: the first block is one-dimensional while the second has dimension $|A|\times|A|$. More generally,  let $(\eta_1\dots\eta_N)$ denote a string of $N$ single-particle quantum numbers that specify an $N$-particle state $|\Psi_{\eta_1\dots\eta_N}\rangle$. The Hamiltonian in Eq.~\eqref{model} commutes with the particle number  $\sum_{j\in\mathbb Z}C^{\dagger}_jC_j$, therefore the reduced density matrix $\rho_A^{(\eta_1\dots\eta_N)}=|\Psi_{\eta_1\dots\eta_N}\rangle\langle\Psi_{\eta_1\dots\eta_N}|$  is a direct sum of operators
\begin{equation}
\label{dec}
\rho_{A}^{(\eta_1\dots\eta_N)}=\oplus_{k=0}^{N}~\rho_{A,[N,k]}^{(\eta_1\dots\eta_N)}.
\end{equation}
The density matrix $\rho_{A,[N,k]}^{(\eta_1,\dots,\eta_N)}$ in Eq.~\eqref{dec} acts in the  $\binom{|A|}{k}$-dimensional subspace of $\mathcal{H}_A$ which contains exactly  $k$ particles. In the following, we also employ the notation $\lambda^{(\eta_1\dots\eta_N)}_{[N,k]}$ for the  eigenvalues of $\rho_{A,[N,k]}^{(\eta_1\dots\eta_N)}$. As shown in Appendix ~\ref{app1}, when $N=1$, the reduced density matrix in Eq.~\eqref{reduced1pt} has only one non-zero eigenvalue in each particle sector and these two eigenvalues are given by
\begin{equation}
 \lambda_{[1,1]}^{(\eta)}=\sum_{r\in A}|\psi_r^{(\eta)}|^2,\quad\lambda_{[1,0]}^{(\eta)}=1-\lambda_{[1,1]}^{(\eta)}.
\end{equation}
Hence, the entanglement entropy, which is computed from Eq.~\eqref{entropy_def}, for the single-particle state given in Eq.~\eqref{sp} reads 
\begin{equation}
\label{ee}
 S^{(\eta)}(A)=-\lambda^{(\eta)}_{[1,1]}\log\lambda_{[1,1]}^{(\eta)}-(1-\lambda_{[1,1]}^{(\eta)})\log(1-\lambda_{[1,1]}^{(\eta)}).
\end{equation}
\section{The $\hbar\rightarrow 0$ limit and the first quantum correction}
\label{sec:classical}
In the continuum limit $a\rightarrow 0$, the eigenvalues of the reduced density matrix are
\begin{equation}
\label{cont_eigen}
 \lambda_{[1,1]}^{(\eta)}\rightarrow\int_{A}dx~|\psi^{(\eta)}(x)|^2,~~\lambda_{[1,0]}^{(\eta)}=1-\lambda_{[1,1]}^{(\eta)}.
\end{equation}
We are now interested in evaluating the $\hbar\rightarrow 0$ limit, hereafter called the classical limit,  of Eq.~\eqref{ee}. As discussed thoroughly in Ref.~\cite{Peres}, depending on the particular observable considered, this procedure  might be ill-defined. However, at least in our  setup, we will  show that all the eigenvalues of the reduced density matrix $\rho_A^{(\eta)}$ converge smoothly to some finite values. In our way of performing the  limit $\hbar\rightarrow 0$ , we assume that the interval $A$ will be inside the classically accessible region bounded by the turning points  $x_l<x_r$ of the classical trajectory. These points are defined as those values for which the classical momentum
\begin{equation} 
\label{cl_mom}
p_{\eta}(x)\equiv\sqrt{2m(E_{\eta}-V(x))}
\end{equation}
vanishes $(p_{\eta}(x_l)=p_{\eta}(x_r)=0)$.  Ignoring for the moment the $O(\hbar)$ corrections that will be discussed  later, the semiclassical wave function in  region $A$ is~\cite{Landau-3}
\begin{equation}
\label{semic}
 \psi_{S}^{(\eta)}(x)=\frac{\mathcal{N}_{\eta}}{\sqrt{p_{\eta}(x)}}\cos\left(\frac{1}{\hbar}\int_{x_l}^{x}dy~p_{\eta}(y)-\frac{\pi}{4}\right),
\end{equation}
 with $\mathcal{N_{\eta}}$ being a normalization constant. In order to replace the wave function $\psi^{(\eta)}$ with its semi-classical approximation $\psi_S^{(\eta)}$, the quantum number $\eta$ must be large, $\eta\gg 1$ (see also Appendix~\ref{app2} for more detail). The normalization constant $\mathcal{N}_{\eta}$ is calculated by neglecting the exponential tails of the semiclassical wave function outside the classically accessible region~\cite{Landau-3} and replacing, inside the integrals, the rapidly oscillating terms with their average, that is
\begin{equation}
 \cos^2\left(\frac{1}{\hbar}\int_{x_l}^{x}dy~p_{\eta}(y)-\frac{\pi}{4}\right)\stackrel{\hbar\rightarrow 0}{\longrightarrow}\frac{1}{2}\,\,\,.
\end{equation}
Therefore one obtains 
\begin{equation}
 \mathcal{N}_{\eta}=2\sqrt{\frac{m}{T_{\eta}}}\,\,\,,
\end{equation}
where $T_{\eta}(E_{\eta})$ is the period of the classical orbit with energy $E_{\eta}=\frac{p^2}{2m}+V(x)$. Hence, in the limit $\hbar\rightarrow 0$, the eigenvalue $\lambda_{[1,1]}^{(\eta)}$ coincides with the probability to observe a classical particle inside interval $A$
\begin{equation}
\label{cl_prob}
 \lambda_{[1,1]}^{(\eta)}\stackrel{\hbar\rightarrow 0}{\longrightarrow} P^{(\eta)}_{\text{cl}}(A)\equiv\frac{2}{T_{\eta}}\int_{A}\frac{d x}{p_{\eta}(x)/m}.
\end{equation}
Analogously, the entanglement entropy calculated in Eq.~\eqref{ee} reduces to 
\begin{equation}
\label{semilimit}
 S^{(\eta)}(A)\stackrel{\hbar\rightarrow 0}{\longrightarrow}-P_{\text{cl}}^{(\eta)}(A)\log P_{\text{cl}}^{(\eta)}(A)-P_{\text{cl}}^{(\eta)}(B)\log P_{\text{cl}}^{(\eta)}(B),
\end{equation}
 with $P_{\text{cl}}^{(\eta)}(B)\equiv 1-P_{\text{cl}}^{(\eta)}(A)$. Notice that  the expression given in Eq.~\eqref{semilimit} is the same as the Shannon entropy~\cite{Shannon} of a classical particle that has probability $P^{(\eta)}_{\text{cl}}(A)$ of being found  inside  interval $A$. For an observer who ignores quantum mechanics but knows that the particle number is conserved, the detection of the particle in  interval $A$ permits one to conclude without hesitation that  interval $B$ must be empty. This is the same classical conditional probability interpretation~\cite{Bell} which could be also employed to explain the anti-correlations measured in a spin-$1/2$ singlet $(|\uparrow\downarrow\rangle-|\downarrow\uparrow)/\sqrt{2}$, as long as the spins  are measured along the {\em same} direction. As is well known, in order to pinpoint the difference between classical and quantum correlations~\cite{Bell2} in a spin system, one should rather measure the two spins along different directions. A similar experiment  however is not easy to reformulate in our model without breaking particle number conservation~\cite{Peres2}.
 
 \vspace{3mm}
 
\noindent\textit{Microcanonical entropy---}
The $\hbar\rightarrow 0$ limit in Eq.~\eqref{semilimit} has also a statistical interpretation in terms of the microcanonical entropy of a classical particle~\cite{Landau-5}, which is observable only within a spatial region $A$. If $\Phi_A$ is the phase space of a classical particle with constant energy $E_{\eta}$ bounded in an interval $A\subset\mathbb R$, its microcanonical entropy is (up to a constant)
\begin{equation}
\label{normalized}
 S^{\text{m}}_{\eta}(A)=-\log[\Gamma_{\eta}(\Phi_A)/T_{\eta})]
\end{equation}
where $\Gamma_{\eta}(\Phi_A)$ is the density of states
\begin{equation}
\label{ds}
 \Gamma_{\eta}(\Phi_A)=\int_{\Phi_A}dp~dx~\delta(p^2/2m+V(x)-E_{\eta}).
\end{equation}
  Given Eq.~\eqref{ds}, Eq.~\eqref{normalized} is the logarithm of the fraction of time that the particle spends inside the interval $A$ during its motion. We can then rewrite Eq.~\eqref{semilimit} as
 \begin{equation}
  S^{(\eta)}(A)\stackrel{\hbar\rightarrow 0}{\longrightarrow}\frac{\Gamma_{\eta}(\Phi_A)}{T_{\eta}}S^{\text{m}}_{\eta}(A)+\frac{\Gamma_{\eta}(\Phi_B)}{T_{\eta}}S_{\eta}^{\text{m}}(B).
 \end{equation}
 Classically, one may  argue that the finite value of the entanglement entropy for $\hbar\rightarrow 0$  is due to the ignorance of the initial condition of the  particle whose motion is confined on a surface (here a curve)  of constant energy. 

\vspace{3mm}
\noindent
\textit{Order $\hbar$ corrections---} We can also easily determine the first quantum correction to the eigenvalue of the reduced density matrix in the one-particle sector, see Eq.~\eqref{cont_eigen}, and therefore to the entanglement entropy in Eq.~\eqref{semilimit}. Let us define~\cite{Landau-3}
\begin{equation}
 \gamma(x) \,=\,\frac{p'}{4p^2}+\frac{1}{8}\int^x dt~\frac{p'^2}{p^3},
\end{equation}
where the $p'$ denotes the derivative of the classical momentum $p$  (see Ref.~\cite{Landau-3} for details). For instance, in the case of the harmonic potential $V(x)=\frac{1}{2}m\omega^2x^2$ and energy $E$ one has
\begin{equation}
 \gamma(y)\,=\,\frac{\omega}{48\sqrt{2} E}\frac{y(y^2-6)}{(1-y^2)^{3/2}},
\end{equation}
with $y=\sqrt{\frac{m\omega^2}{4E}} x$.
Up to a normalization constant, the semiclassical wavefunction which includes also  $O(\hbar)$ corrections is  obtained by replacing Eq.~\eqref{semic} with~\cite{Landau-3} 
\begin{equation}
\label{semic_1}
 \psi_S^{(\eta)}\rightarrow \psi_S^{(\eta)}(1-i\hbar \gamma(x)).
\end{equation}
One can normalize the new wave function in Eq.~\eqref{semic_1} by ignoring as before the exponential tails outside the classically accessible region and eventually derive  the first quantum correction to Eq.~\eqref{cl_prob} as
\begin{multline}
 \lambda_1^{(\eta)}=P_{\text{cl}}^{(\eta)}(A)+\frac{2\hbar^2}{T_{\eta}}\left(\int_A dx~\frac{\sigma_2^2}{p_{\eta}}-P_{\text{cl}}(A)\int_{x_l}^{x_r} dx~ \frac{\sigma_2^2}{p_{\eta}}\right)\\+O(\hbar^3).
\end{multline}

\section{Classical limit of two-particle state entanglement entropy}
\label{sec:tp}
Let us now examine the same problem but for a fermionic two-particle eigenstate of the Hamiltonian \eqref{model}, namely
\begin{equation}
\label{2ptstate}
 |\Psi_{\eta\beta}\rangle=\sum_{l<m\in\mathbb Z} M_{lm}^{(\eta \beta)}|1_l 1_m\rangle,
\end{equation}
with $M_{lm}^{(\eta \beta)}=\psi_l^{(\eta)}\psi_{m}^{(\beta)}-\psi_{m}^{(\eta)}\psi_{l}^{(\beta)}$ and $\eta\not=\beta$. The two-particle state satisfies the anticommutation relation
\begin{equation}
 |1_l1_m\rangle=-|1_m1_l\rangle.
\end{equation}
Therefore an orthonormal basis for the two-particle sector of $\mathcal{H}$ is given by the vectors $|1_l1_m\rangle$ with $l<m$. The state in Eq.~\eqref{2ptstate} is properly normalized, $\langle\Psi_{\eta\beta}|\Psi_{\eta\beta}\rangle=1$. By repeating steps similar to those that led us to Eq.~\eqref{reduced1pt}, we can determine the reduced density matrix of region $A$ for this case 
\begin{multline}
\label{rho_red}
 \rho_A^{(\eta\beta)}\,=\, \sum_{\substack{l<m\in A \\ l'<m'\in A}}M_{lm}^{(\eta \beta)} [M_{l'm'}^{(\eta \beta)}]^{*}|1_{l}1_{m}\rangle\langle 1_{m'}1_{l'}|+\\+\sum_{l<m\in B} |M^{(\eta \beta)}_{lm}|^{2}\Pi_{0}^{A} 
 +\sum_{\substack{l\in B\\ m,m'\in A}}M_{lm}^{(\eta \beta)} [M^{(\eta \beta)}_{lm'}]^{*}|1_m\rangle\langle 1_{m'}|. \nonumber
\end{multline}
As a matrix acting on $\mathcal{H}_A$, $\rho_A^{(\eta \beta)}$ is a direct sum of three orthogonal projectors into the two-, zero- and one-particle sectors of the Hilbert space.
Let us now consider  the $\hbar\rightarrow 0$ limit of its eigenvalues.
The reduced density matrix in the two-particle sector has rank one (see Appendix~\ref{app1}), and its only non-vanishing eigenvalue equals the trace, therefore
\begin{equation}
\label{eigen_20pt}
 \lambda_{[2,2]}^{(\eta \beta)}=\sum_{l,m\in A}|\psi_l^{(\eta)}|^2|\psi_{m}^{(\beta)}|^2-\left|\sum_{l\in A}[\psi_{l}^{(\eta)}]^{*}\psi_{l}^{(\beta)}\right|^2.
\end{equation}
As expected, $\lambda_{[2,0]}^{(\eta \beta)}$ and $\lambda_{[2,2]}^{(\eta \beta)}$ are related by the exchange of the bipartition indices $A$ and $B$. 
In the continuum limit,  Eq.~\eqref{eigen_20pt} reduces to 
\begin{equation}
\label{cont_l}
 \lambda_{[2,2]}^{(\eta \beta)}\rightarrow\det_{\eta,\beta}\int_{A} dx~[\psi^{(\eta)}(x)]^{*}\psi^{(\beta)}(x),
\end{equation}
with $\psi^{(\eta)}(x) $ and $\psi^{(\beta)}(x)$ being the single particle Schr\"odinger wave functions with eigenvalues $E_{\eta,\beta}$, $(\eta\not=\beta)$. As in Sec.~\ref{sec:classical}, we  take $A$  within the region classically accessible to both particles with energies $E_{\eta}$ and $E_{\beta}$. 
 The matrix in Eq.~\eqref{cont_l} becomes then  diagonal for $\hbar\rightarrow 0$. Indeed, by applying the stationary phase approximation  (see Appendix~\ref{app2}), one has
\begin{equation}
\label{semi_overlap}
 \int_{A}dx~[\psi_{S}^{(\eta)}(x)]^{*}\psi_{S}^{(\beta)}(x)\stackrel{\hbar\rightarrow 0}{\longrightarrow}\delta_{\eta,\beta}P_{\text{cl}}^{(\eta)}(A);
\end{equation}
and therefore we obtain $\lambda_{[2,2]}^{(\eta \beta)}\stackrel{\hbar\rightarrow 0}{\longrightarrow}P_{\text{cl}}^{(\eta)}(A)P_{\text{cl}}^{(\beta)}(A)$. The eigenvalues of the reduced density matrix in the one-particle sector can be calculated similarly. By using again Eq.~\eqref{semi_overlap}, in the continuum limit we have 
\begin{equation}
\label{fermion1p}
 \sum_{\substack{l\in B\\ m,m'\in A}}M_{lm}^{(\eta \beta)} [M^{(\eta \beta)}_{l'm'}]^{*}\stackrel{\hbar\rightarrow 0}{\longrightarrow} K_S^{(\eta \beta)}(y,y') 
\end{equation}
where 
\begin{eqnarray}
K_S^{(\eta \beta)}(y,y') \equiv&& \psi_S^{(\eta)}(y)[\psi_S^{(\eta)}(y')]^{*}P_{\text{cl}}^{(\beta)}(B)+\nonumber\\
&& +\psi_S^{(\beta)}(y)[\psi_S^{(\beta)}(y')]^{*} P_{\text{cl}}^{(\eta)}(B),
\end{eqnarray}
for $y,y'\in A$.
 The kernel $K^{(\eta \beta)}_S(y,y')$, acting on the one-particle semiclassical wave functions, has only two non-zero eigenvalues: one is $P^{(\eta)}_{\text{cl}}(A)P_{\text{cl}}^{(\beta)}(B)$ and  the other is $P_{\text{cl}}^{(\beta)}(A) P^{(\eta)}_{\text{cl}}(A)$. By taking into account Eq.~\eqref{eigen_20pt}, we conclude that the classical limit of the eigenvalues of the reduced density matrix $\rho_A^{(\eta \beta)}$ is
\begin{align}
\label{eigen_dm} 
& \lambda_{[2,0]}^{(\eta \beta)}\stackrel{\hbar\rightarrow 0}{\longrightarrow}P_{\text{cl}}^{(\eta)}(B)P_{\text{cl}}^{(\beta)}(B) \,\,\,
 \\
 \label{eigen_dm2} 
&\lambda_{[2,1]}^{(\eta \beta)}\stackrel{\hbar\rightarrow 0}{\longrightarrow}\{P_{\text{cl}}^{(\eta)}(A)P_{\text{cl}}^{(\beta)}(B),~P_{\text{cl}}^{(\beta)}(A)P_{\text{cl}}^{(\eta)}(B)\},\\
\label{eigen_dm3} 
& \lambda_{[2,2]}^{(\eta \beta)}\stackrel{\hbar\rightarrow 0}{\longrightarrow}P_{\text{cl}}^{(\eta)}(A)P_{\text{cl}}^{(\beta)}(A). 
\end{align}
Notice that their sum is $1$ and moreover they have a simple  combinatorial interpretation, already anticipated in Sec.~\ref{sec:classical}. The eigenvalues in Eqs.~(\ref{eigen_dm})-(\ref{eigen_dm3}) represent the probability to observe zero, one or two particles of different colors $\eta$ and $\beta$ into an interval  $A$. The entanglement entropy $S^{(\eta \beta)}(A)=-\text{Tr}[\rho_A^{(\eta \beta)}\log\rho_{A}^{(\eta \beta)}]$ converges for $\hbar\rightarrow 0$ to the Shannon entropy~\cite{Shannon} of such probability distribution
\begin{multline}
\label{s2p}
S^{(\eta \beta)}(A)\stackrel{\hbar\rightarrow 0}{\longrightarrow}-\sum_{r=\eta,\beta}\left[P_{\text{cl}}^{(r)}(A)\log P^{(r)}_{\text{cl}}(A)\right.\\
\left.+(1-P_{\text{cl}}^{(r)}(A))\log (1-P^{(r)}_{\text{cl}}(A))\right] 
\end{multline}
 
\vspace{3mm}
\noindent
\textit{Identical quantum numbers---}It is interesting to examine more closely the case $\eta=\beta$, which requires the particles to be bosonic. The normalized two-particle state is now
\begin{equation}
\label{bosons}
 |\Psi\rangle=\sqrt{2}\sum_{l<m\in\mathbb Z}\psi^{(\eta)}_l\psi_{m}^{(\eta)}|1_l 1_m\rangle+\sum_{l\in\mathbb Z}[\psi_l^{(\eta)}]^2|2_l\rangle,
\end{equation}
where $|2_l\rangle=\frac{1}{\sqrt{2}}(C^{\dagger}_l)^2|\Omega\rangle$. However, in the continuum limit (see Eq.~\eqref{cont_lim}), the second term in Eq.~\eqref{bosons} is $O(a)$ and drops;  therefore all the steps previously done for fermions can be repeated but with a crucial difference. The kernel $K_S(y,y')$ in Eq.\eqref{fermion1p} is now replaced by
\begin{equation}
\label{kernel_b}
 K_S^{(\eta\eta)}(y,y)\,=\,2\psi_S^{(\eta)}(y)[\psi_S^{(\eta)}(y')]^{*}P_{\text{cl}}^{(\eta)}(B),~y,y'\in A,
\end{equation}
and has only one non-zero eigenvalue given by $\lambda_{[1,1]}^{(\eta\eta)}=2P_{\text{cl}}^{(\eta)}(A)P_{\text{cl}}^{(\eta)}(A)$. The eigenvalues $\lambda_{[2,k]}^{(\eta\eta)}$ of the reduced density matrix are now probabilities of occurrences of $k=0,1,2$ successes in $N=2$ Bernoulli trials (coin tossing). The entanglement entropy for $\hbar\rightarrow 0$ is then the Shannon entropy of a binomial distribution $B(N,p)$ with the following parameters: $N=2$, the number of trials, and $p=P_{\text{cl}}^{(\eta)}(A)$, the probability of success in each trial.  Notice that this result cannot be obtained by substituting $\eta=\beta$ in Eq.~\eqref{s2p}.

\section{Classical limit for an arbitrary number of particles}
\label{sec:gen}
 A generalization of the results  derived in Sec.~\ref{sec:tp} is also easy to obtain  for arbitrary multiparticle fermionic states.  An $N$-particle eigenstate of  Eq.~\eqref{model} is given in this case by 
\begin{equation}
|\Psi_{\eta_1\dots\eta_N}\rangle=\sum_{l_1<\dots<l_N\in\mathbb Z} M_{l_1\dots l_N}^{(\eta_1\dots \eta_N)}|1_{l_1}\dots 1_{l_N}\rangle, 
\end{equation}
with $M_{l_1\dots l_N}^{(\eta_1\dots\eta_N)}=\det[\psi^{(\eta_i)}_{l_{j}}]_{i,j=1\dots N}$. The reduced density matrix acting on the subspace of $\mathcal{H}_A$ with exactly $k$ particles, see Eq.~\eqref{dec},  can be written as
\begin{eqnarray}
\label{mpts}
&& \rho_{A,[N,k]}^{(\eta_1\dots\eta_N)}=\sum_{\substack{m_1<\dots <m_k \in A\\ m_1'<\dots< m_k' \in A}}~  \sum_{l_1<\dots<l_{N-k}\in B}M_{l_1\dots l_{N-k}m_1\dots m_k}^{(\eta_1\dots\eta_N)} \nonumber \\
&& \times [M_{l_1\dots l_{N-k}m_1'\dots m_k'}^{(\eta_1\dots\eta_N)}]^{*} 
  \times \, |1_{m_1}\dots 1_{m_k}\rangle\langle 1_{m_1'}\dots 1_{m_k'}|.  
\end{eqnarray}
 In order to calculate the eigenvalues of $\rho_{A,[N,k]}^{(\eta_1\dots\eta_n)}$ in the limit $\hbar\rightarrow 0$, we proceed as follows.
First, the product of the two determinants in Eq.~\eqref{mpts}, is expanded over permutations as
\begin{multline}
\label{Leib} 
M_{l_1\dots l_{N-k}m_1\dots m_k}^{(\eta_1\dots\eta_N)}[M_{l_1\dots l_{N-k}m_1'\dots m_k'}^{(\eta_1\dots\eta_N)}]^{*}=\\
 \sum_{\sigma,\tau\in S_N}(-1)^{\sigma+\tau} \psi_{l_1}^{(\eta_{\sigma(1)})}\dots\psi_{l_{N-k}}^{(\eta_{\sigma(N-k)})}\psi_{m_1}^{(\eta_{\sigma(N-k+1)})}\dots\psi_{m_k}^{(\eta_{\sigma(N)})} \\
\times [\psi_{l_1}^{(\eta_{\tau(1)})}]^{*}\dots[\psi_{l_{N-k}}^{(\eta_{\tau(N-k)})}]^{*}[\psi_{m_1'}^{(\eta_{\tau(N-k+1)})}]^{*}\dots[\psi_{m_k'}^{(\eta_{\tau(N)})}]^{*}.
 \end{multline}
Then, we observe that, by Eq.~\eqref{semi_overlap}, when summing Eq.~\eqref{Leib} over $l_1,\dots,l_{N-k}$,  the limit $\hbar\rightarrow 0$ selects only the permutations with $\sigma(1)=\tau(1),\dots,\sigma(N-k)=\tau(N-k)$.
Each of these identifications of the $N-k$ quantum numbers can be performed in $\binom{N}{k}$ distinct  ways and for a given choice of the first $N-k$ indices there are  $(N-k)!(k!)^2$ terms in the summation in Eq.~\eqref{Leib}.  For instance, if we  choose  $\sigma(1)=\tau(1),\dots,\sigma(N-k)=\tau(N-k)$  within the set $\{k+1,\dots,N\}$, these terms will be factorized and are of the form
\begin{multline}
\label{fact}
\left(|\psi_{l_1}^{(\eta_{k+1})}|^2\dots...|\psi_{l_{N-k}}^{(\eta_{N})}|^2+\text{perm.}\right)\\
\times \sum_{\sigma,\tau\in S_{k}}(-1)^{\sigma+\tau}\psi_{m_1}^{(\eta_{\sigma(1)})}\dots\psi_{m_k}^{(\eta_{\sigma(k)})}[\psi_{m_1'}^{(\eta_{\tau(1)})}]^{*}\dots[\psi_{m_k'}^{(\eta_{\tau(k)})}]^{*}. 
\end{multline}
By substituting back Eq.~\eqref{fact} into Eq.~\eqref{mpts} and summing over $l_1,\dots, l_{N-k}$,
we deduce that the non-vanishing contributions to the classical limit  in the $k$-particle sector of $\mathcal{H}_A$ are
\begin{equation}
\label{factorization}
 \rho_{A,[N,k]}^{(\eta_1\dots\eta_N)}\rightarrow\sum_{\mathcal{S}=\{i_1,\dots,i_{k}\}}\left(\prod_{j\not\in\mathcal{S}}\sum_{l\in B}|\psi_{l}^{(\eta_j)}|^2\right)\rho_{A,[k,k]}^{(\eta_{i_1}\dots\eta_{i_k})},
\end{equation}
where  $\mathcal{S}$ is a $k$-tuple of indices in the set $\{1,\dots, N\}$. By applying the Cauchy-Binet theorem (see  Appendix~\ref{app1}), we can also find
\begin{multline}
 \label{Binet}
  \rho_{A,[k,k]}^{(\eta_1\dots\eta_k)}\circ\rho_{A,[k,k]}^{(\beta_1\dots\beta_k)}=\left(\det_{\eta,\beta}\sum_{l\in A}\psi_{l}^{(\eta)}[\psi^{(\beta)}_l]^{*}\right)
   \times  \\
   \sum_{\substack{m_1<\dots<m_k\\ m_1'<\dots<m_k'}}
    \det(\psi^{(\eta_i)}_{m_j})\det([\psi^{(\beta_i)}_{m_j'}]^{*})\times\\  |1_{m_1}\dots 1_{m_k}\rangle\langle 1_{m_k'}\dots 1_{m_1'}|,
\end{multline}
which shows, again recalling Eq.~\eqref{semi_overlap}, that all the density matrices in Eq.~\eqref{factorization} commute when $\hbar\rightarrow 0$. Moreover, the operators $\rho_{A,[k,k]}^{\eta_1\dots\eta_k}$ have a unique non-zero eigenvalue given in the continuum limit by  Eq.~\eqref{eq_eigen_k}
\begin{equation}
\label{cont}
 \lambda_{[k,k]}^{(\eta_1\dots \eta_k)}=\det_{i,j}\int_A dx~\psi^{(\eta_i)}(x)[\psi^{(\eta_j)}(x)]^{*}.
\end{equation}
The eigenvectors relative to the eigenvalues in Eq.~\eqref{cont} are  orthogonal in the classical limit for different sets of quantum numbers $\{\eta_1,\dots,\eta_k\}$-- [see Eqs.~\eqref{eigen_prod} and~\eqref{semi_overlap}]. Hence, we can conclude that the density matrix  $\rho_{A,[N,k]}^{(\eta_1\dots\eta_N)}$ has, for $\hbar\rightarrow 0$, $\binom{N}{k}$ distinct eigenvalues
\begin{equation}
\label{eigen_multi}
 \lambda_{[N,k]}^{(\eta_1\dots\eta_N)}\stackrel{\hbar\rightarrow 0}{\rightarrow}\prod_{j\not\in S}P^{(\eta_j)}_{\text{cl}}(B)\prod_{j\in\mathcal{S}}P_{\text{cl}}^{(\eta_j)}(A),
\end{equation}
labeled by the $k$-tuples $\mathcal{S}$ of indices in the set $\{1,\dots, N\}$. Once again (see for instance Sec.~\ref{sec:tp}), the eigenvalues $\lambda^{(\eta_1\dots\eta_N)}_{[N,k]}$ in the classical limit have a simple combinatorial interpretation: they are the probabilities associated with the possible arrangements of  $k$ out of $N$ colored particles in the interval $A$. For $\hbar\rightarrow 0$, the total number of non-zero eigenvalues of $\rho_{A}^{(\eta_1\dots\eta_N)}$ is $2^N$ and the entanglement entropy converges to their Shannon entropy
\begin{eqnarray}
\label{sp_entropy}
S^{(\eta_1\dots\eta_N)}(A)\stackrel{\hbar\rightarrow 0}{\longrightarrow}-\sum_{r=1}^N[P_{\text{cl}}^{(\eta_r)}(A)\log P^{(\eta_r)}_{\text{cl}}(A)
+ \\
+ (1-P_{\text{cl}}^{(\eta_r)}(A))\log (1-P^{(\eta_r)}_{\text{cl}}(A))]. \nonumber
\end{eqnarray}
Notice that $S^{(\eta_1\dots\eta_N)}(A)\leq N\log 2$ with the bound saturated for $P_{\text{cl}}^{(\eta_r)}(A)=1/2$,  $\forall r=1,\dots N$. For a system of $N$ fermions on a ring of length $L$, one has $P_{\text{cl}}(A)=|A|/L$ and Eq.~\eqref{sp_entropy} coincides  with the universal part of the quasi-particle excited state entanglement entropy calculated with field theory techniques in  the limit of large and distinct momenta in Refs.~\cite{Castro, Castro2}.
More precisely, Refs.~~\cite{Castro, Castro2} obtained Eq.~\eqref{sp_entropy} for a gas of $N$ bosonic particles with different quantum numbers. The  agreement with our fermionic calculation is however not surprising since for free particles on a ring, either bosons or fermions, the limit of large momenta is, by the correspondence principle, the classical limit defined here.

We conclude this Section with two remarks.
The entanglement entropy in the ground state of a Fermi gas in an external potential has been discussed in the limits $\hbar\rightarrow 0$ and $N\rightarrow\infty$ by the authors of Ref.~\cite{Dubail}. When $N\hbar$ is finite,  this quantity can be calculated with a field theoretical approach, see also~Ref.~\cite{Brun} for additional details on this way of taking the  semiclassical limit.

Finally, we observe  that Eq.~\eqref{cont} can be  interpreted as the Emptiness Formation Probability~\cite{KBI} of region $B$.  Indeed Eq.~\eqref{cont}  can be rewritten as
\begin{equation}
\label{emptiness}
 \mathcal{E}(B)=\det_{\eta,\beta}\left[\delta_{\eta,\beta}-\int_{B}dx~\psi^{(\eta)}(x)[\psi^{(\beta)}(x)]^{*}\right]\,\,\,,
\end{equation}
and, by using $\det(1-K)=-\sum_{p\geq 1}\text{Tr}(K^p)/p$, it can be recast in a Fredholm determinant form with a Christoffel-Darboux kernel
\begin{eqnarray}
\label{CrD}
&& \mathcal{E}(B)=\det(1-K(x,y))|_{x,y\in B},\\
&&K(x,y)=\sum_{\eta=1}^{N}[\psi^{(\eta)}(x)]^{*}\psi^{(\eta)}(y).\nonumber
\end{eqnarray}
Eq.~\eqref{CrD} is a well known formula in the random matrix literature, see for instance ~\cite{DML}.

\section{Conclusions}
\label{sec:conc}
In this paper we  investigated the classical limit of the eigenvalues of the reduced density matrix of a one-dimensional fermionic quantum gas in an external potential.  We showed that the eigenvalues of the reduced density matrix $\rho_A$ of a spatial interval $A$ are finite for $\hbar\rightarrow 0$.  They can be interpreted classically in terms of probabilities of distinct arrangements of $k$ particles $(k=1,\dots,N)$ with $N$ different colors into two boxes. Moreover, the entanglement entropy of the subsystem $A$ reduces to the Shannon entropy of $N$ bits. A similar conclusion can be also found in Refs.~\cite{Castro, Castro2} as a result of a field theoretical calculation for free bosons on a ring in the limit of large and distinct momenta. Our analytic derivation however does not rely on field theoretical tools---such as twist fields or replicas---and  generalizes  the results in~ Refs.~\cite{Castro, Castro2} to  fermions in an arbitrary external potential. It also suggests that the  universal part~\cite{Castro, Castro2, Piz, Raja_Z} of the quasi-particle excited state entanglement entropy has a classical origin. 
For $N=2$, we  analyzed the possibility that the quantum particles have the same quantum numbers and therefore are bosons.  In this case, it turns out that the classical limit of the entanglement entropy of a spatial region $A$ coincides with the Shannon entropy of a binomial distribution of two Bernoulli trials (coin tossing).

It would be interesting to generalize this calculation, as done in in Sec.~\ref{sec:gen}, to $N$ identical quantum numbers. If the eigenvalues of the reduced density matrix  in the $k$-particle sector of the Hilbert space still coincide with the probabilities of $k$ successes in $N$ independent Bernoulli trials, the entanglement entropy   will converge to the Shannon entropy of a binomial distribution, see also~ Ref.~\cite{Castro2}. Curiously, for large $N$, the latter also scales logarithmically with $N$, as found for instance in critical bosonic and fermionic one-dimensional systems at zero temperature~\cite{Korepin, CC}.

Our example suggests that, even for a pure state, the entanglement entropy might be finite for $\hbar\rightarrow 0$ and therefore, in this case, must admit a consistent classical probability interpretation.  Following~Ref.~\cite{Bell2}, in order to pinpoint unambiguously the non-classical behavior of the correlations in a quantum superposition one should  try to set up a concrete experiment. For instance, Peres in Ref.~\cite{Peres2} proposed the one where the measurement apparatus could change the particle number of the initial quantum state but, as we have already mentioned, this violation is not possible in our simple model.

Finally, we mention that other possible extensions of the work are represented by the study of the classical limit of the mutual information or the negativity~\cite{Peres3, Vidal} analyzed for free fermionic theories in~Refs.~\cite{Tonni, Tonni2, Ryu1, Ryu2}.

\acknowledgements
We are grateful to Filiberto Ares, Leonardo Banchi, Pasquale Calabrese, John Cardy, Filippo Colomo and Erik Tonni for discussions.  JV would like to thank the INFN of Florence and in particular Andrea Cappelli for the kind hospitality. He  acknowledges partial support by the  the CNPq (grant number 306209/2019-5) and the Simons Foundation (Grant
Number 884966, AF. GM acknowledges the grant Prin $2017$-FISI.

\onecolumngrid
\appendix
\section{Linear Algebra tools}
\label{app1}
\subsection*{A simple proposition}
Given any vector $Q\in\mathbb C^{N}$, and an orthonormal basis $|v_k\rangle$ for a $N$-dimensional Hilbert space on $\mathbb{C}$, the linear operator
\begin{equation}
\label{simprop}
 \rho(Q)=\sum_{k,k'=1}^{N}Q_{k}[Q_{k'}]^{*}|v_k\rangle\langle v_{k'}|
\end{equation}
has rank one. 

\vspace{3mm}
\noindent
\textit{Proof.} Let us apply $\rho$ to a vector $|u(R)\rangle=\sum_{k=1}^{N} R_{k}|v_k\rangle$, we get
\begin{equation}
 \rho(Q)|u(R)\rangle=(Q^{\dagger}R)\sum_{k=1}^{N}Q_{k}|v_k\rangle.
\end{equation}
By taking $R$ in the orthogonal complement of $Q$, which has dimension $N-1$, we obtain $\rho(Q) |u(R)\rangle=O$, while by selecting $R$ parallel to $Q$ we have $\rho(Q)|u(R)\rangle=|Q|^2|u(R)\rangle$. This  proves that $\rho(Q)$ has $N-1$ vanishing eigenvalues and one positive eigenvalue equal to $|Q|^2.\hfill\blacksquare$

The same conclusion also follows from the fact that $\rho(Q)$ is a projector on the state $\sum_{k=1}^N Q_k|v_k\rangle$. Finally, if $|u(Q)\rangle$ and $|u(Q')\rangle$ are eigenvectors of $\rho(Q)$ and $\rho(Q')$ one also has
\begin{equation}
\label{sp2}
 \langle u(Q)|u(Q')\rangle=Q^{\dagger}Q'.
\end{equation}

\subsection*{Proof of Eq.~\eqref{Binet} (Cauchy-Binet theorem)}
From the definitions given in the main text
\begin{multline}
\rho_{A,[k,k]}^{(\eta_1\dots\eta_k)}\circ\rho_{A,[k,k]}^{(\beta_1\dots\beta_k)}=\sum_{\substack{n_1<\dots<n_k \\ n_1'<\dots<n_k'}}\sum_{\substack{m_1<\dots<m_k\\ m_1'<\dots<m_k'}}\sum_{\substack{\sigma,\tau\in S_k\\ \lambda,\mu \in S_k}} 
 (-1)^{\sigma+\tau}(-1)^{\lambda+\mu}\psi_{m_1}^{(\eta_{\sigma(1)})}\dots\psi_{m_k}^{(\eta_{\sigma(k)})}[\psi_{m_1'}^{(\beta_{\tau(1)})}]^{*}\dots[\psi_{m_k'}^{(\beta_{\tau(k)})}]^{*}\times\\
 \psi_{n_1}^{(\eta_{\lambda(1)})}\dots\psi_{n_k}^{(\eta_{\lambda(k)})}[\psi_{n_1'}^{(\beta_{\mu(1)})}]^{*}\dots[\psi_{n_k'}^{(\beta_{\mu(k)})}]^{*}
|1_{m_1}\dots 1_{m_k}\rangle\langle 1_{m_k'}\dots 1_{m_1'}|1_{n_1}\dots 1_{n_k}\rangle\langle 1_{n_k'}\dots 1_{n_1'}|.
\end{multline}
The scalar product on the second line gives $\prod_{i=1}^k\delta_{m_i',n_i}$, since the basis is orthonormal, we then obtain
\begin{multline}
\label{dmcomm}
\rho_{A,[k,k]}^{(\eta_1\dots\eta_k)}\circ\rho_{A,[k,k]}^{(\beta_1\dots\beta_k)}=
\sum_{\substack{m_1<\dots<m_k \\ m_1'<\dots<m_k'}}\det(\psi^{(\eta_i)}_{m_j})\det([\psi^{(\beta_i)}_{m_j'}]^{*})
\left(\sum_{n_1<\dots<n_k}\sum_{\tau,\lambda\in S_k}(-1)^{\tau+\lambda}
\prod_{j=1}^{k}\psi^{(\beta_{\tau(j)})}_{n_j}[\psi^{(\eta_{\lambda(j)})}_{n_j}]^{*}\right)\times \\
|1_{m_1}\dots 1_{m_k}\rangle\langle 1_{m_k'}\dots 1_{m_1'}|.
\end{multline}
The indices $n_i\in A$ and therefore run on $|A|$ possible values. Consider now the $|A|\times k$ rectangular matrix $X_{n j}=\psi^{(\eta_j)}_{n}$ and the $k\times |A|$ rectangular matrix $Y_{jn}=[\psi^{(\beta_j)}_n]^{*}$. Let $S=\{n_1,\dots,n_k\}$ a $k$-tuple of rows of $X$ or columns of $Y$, the Cauchy-Binet theorem states that (see Ref.~ \cite{CB_paper})
\begin{equation}
\label{CB}
 \det (X^{T}Y)=\sum_{\mathcal{S}}\det X_{\mathcal{S}}\det Y_{\mathcal{S}}.
\end{equation}
We recognize then the prefactor in the first line of Eq.~\eqref{dmcomm} as the right hand side of Eq.~\eqref{CB} and conclude that
\begin{multline}
\label{eq_Binet}
\rho_{A,[k,k]}^{(\eta_1\dots\eta_k)}\circ\rho_{A,[k,k]}^{(\beta_1\dots\beta_k)}=\left(\det_{\eta,\beta}\sum_{l\in A}\psi_{l}^{(\eta)}[\psi^{(\beta)}_l]^{*}\right)
 \sum_{\substack{m_1<\dots<m_k\\ m_1'<\dots<m_k'}}\det(\psi^{(\eta_i)}_{m_j})\det([\psi^{(\beta_i)}_{m_j'}]^{*})|1_{m_1}\dots 1_{m_k}\rangle\langle 1_{m_k'}\dots 1_{m_1'}|.
\end{multline}
This is the formula given in the Eq.~\eqref{Binet} of the main text.
Notice that if $\eta_1=\beta_1,\dots,\eta_{k}=\beta_k$, Eq.~\eqref{eq_Binet} implies that $(\rho_{A,[k,k]}^{(\eta_1\dots\eta_k)})^{2}=\lambda_{k}^{(\eta_1\dots\eta_k)}\rho_{A,[k,k]}^{(\eta_1\dots\eta_k)}$ with
\begin{equation}
\label{eq_eigen_k}
 \lambda_{[k,k]}^{(\eta_1\dots\eta_k)}=\det_{i,j}\sum_{l\in A}\psi^{(\eta_i)}_l[\psi^{(\eta_j)}_l]^*.
\end{equation}
From Eq.~\eqref{simprop} it follows that the operator $\rho_{A,[k,k]}^{(\eta_1\dots\eta_k)}$ has rank one and therefore $\lambda_{[k,k]}^{(\eta_1\dots\eta_k)}$ is its only non-zero eigenvalue. Let $|u^{(\eta_1\dots\eta_k)}\rangle$ be the corresponding eigenvector, then Eq.~\eqref{sp2} and the Cauchy-Binet theorem also imply that
\begin{equation}
\label{eigen_prod}
 \langle u^{(\eta_1\dots\eta_k)}|u^{(\beta_1\dots\beta_k)}\rangle=\left(\det_{\eta,\beta}\sum_{l\in A}\psi_{l}^{(\eta)}[\psi^{(\beta)}_l]^{*}\right).
\end{equation}

\section{$\hbar\rightarrow 0$ limit of the overlaps (Eq.~\eqref{semi_overlap})}
\label{app2}
The overlaps between semiclassical wave functions were discussed in the pedagogical note~\cite{Wheeler}. In the limit $\hbar\rightarrow 0$, they can be evaluated by stationary phase approximation. By substituting Eq.~\eqref{semic} into Eq.~\eqref{semi_overlap} one finds
\begin{equation}
\label{int}
\int_{A} dx~[\psi_S^{(\eta)}(x)]^{*}\psi_S^{(\beta)}(x)= T_{\eta}T_{\beta}\, \int_{A} dx~\frac{e^{\frac{i}{\hbar}g_{\eta\beta}(x)}-ie^{\frac{i}{\hbar}f_{\eta\beta}(x)}}{\sqrt{p_{\eta}(x)p_{\beta}(x)}}+\text{cc}, 
\end{equation}
where $f_{\eta\beta}(x)=\mathcal{A}_{\eta}(x)+\mathcal{A}_{\beta}(x)$, $g_{\eta\beta}(x)=\mathcal{A}_{\eta}(x)-\mathcal{A}_{\beta}(x)$ and $\mathcal{A}_{\eta}(x)$ is the modified action 
 $\mathcal{A}_{\eta}(x)=\int_{x_l}^{x} dy~p_{\eta}(y)$ and  $p_{\eta}(y)>0$ is the classical momentum in Eq.~\eqref{cl_mom}.
In the limit $\hbar\rightarrow 0$, the asymptotics of the integral in Eq.~\eqref{int} is dominated by the contributions of stationary points such that $g_{\eta\beta}'(x_s)=0$ (since $f'_{\eta\beta}(x)\not=0$ for $x\in A$) and of points located at the boundary of the integration domain $A$. The latter were ignored in Ref.~\cite{Wheeler} since they were producing subleading terms. If $x_s$ is a stationary points of $g_{\eta\beta}$, then $p_{\eta}(x_s)=p_{\beta}(x_s)$, however for $E_{\eta}\not=E_{\beta}$, classical trajectories with the same potential cannot cross in phase space. We conclude therefore that  $g'_{\eta\beta}(x)\not=0$ for $x\in A$. It remains to analyze the boundary contributions to the asymptotics. Let us consider for instance
\begin{equation}
\label{int_check}
 I=\int_{A} dx\frac{-ie^{\frac{i}{\hbar} f_{\eta\beta}(x)}}{\sqrt{p_{\eta}(x)p_{\beta}(x)}}.
\end{equation}
By integrating by parts we get
\begin{equation}
\label{asympt}
I=-\left.\frac{\hbar e^{\frac{i}{\hbar}f_{\eta\beta}(x)}}{(p_{\eta}(x)+p_{\beta}(x))\sqrt{p_{\eta}(x)p_{\beta}(x)}}\right|_{\partial A}+\hbar\int_{A}dx~\frac{d}{dx}\left(\frac{1}{(p_{\eta}(x)+p_{\beta}(x))\sqrt{p_{\eta}(x)p_{\beta}(x)}}\right) e^{\frac{i}{\hbar}f_{\eta\beta}(x)},\nonumber
\end{equation}
which shows that the boundary contribution to $I$ is $O(\hbar)$ and vanishes for $\hbar\rightarrow 0$. As an example of the quality of the asymptotics in Eq.~\eqref{asympt}, we calculated Eq.~\eqref{int_check} for the harmonic potential $V(x)=\frac{1}{2}x^2$ taking $\eta=10$ and $\beta=20$ at $\hbar=1$. One should not be puzzled by the choice $\hbar=1$. The classical limit is approached whenever $\mathcal{A}_{\eta,\beta}\gg \hbar$ (for $x\in A$) and therefore the stationary phase approximation remains also valid for $\hbar=O(1)$ as long as $\eta,\beta\gg 1$.   For an interval $A=[-1,1]$, we obtain by numerical integration of Eq.~\eqref{int_check} $I= 0.03446-0.00437i$ while  the first term in Eq.~\eqref{asympt} is evaluated as $0.03445-0.00437i$ in excellent agreement.


\begin{thebibliography}{99}
\bibitem{Foot} The bosonic case can be discussed as well, see Sec.~\ref{sec:tp}.
\end{thebibliography}
\end{document}